\documentstyle[12pt]{article}

\begin{document}
\begin{center}
{\Large \bf The third order effect for the electromagnetic wave
in the frame moving transverse to the wave}
\bigskip

{\large D.L.~Khokhlov}
\smallskip

{\it Sumy State University, R.-Korsakov St. 2, \\
Sumy 40007, Ukraine\\
E-mail: dlkhokhl@rambler.ru}
\end{center}

\begin{abstract}
It is considered the electromagnetic wave in the frame moving
transverse to the wave. Due to the
velocity of the frame with respect to a preferred reference frame
the oscillations of phase arise. The oscillations of phase yield
the third order effect which can be seen in the Michelson-Morley
experiment. The effect due to the velocity of the earth with
respect to CMB is in agreement with the anisotropy determined in
the Michelson-Morley experiment.

\end{abstract}

Consider the plane monochromatic electromagnetic wave. Let the
wave vector $k$ be directed along the axis $x$, the electric field
$E$ along the axis $y$, the magnetic field $H$ along the axis $z$.
Suppose that the invariance of the velocity of light holds true.
Let the
emitter and receiver of the electromagnetic wave be situated in
the frame moving with the velocity $v$ with respect to a
preferred reference frame. We shall take the frame of the cosmic
microwave background radiation (CMB) as a preferred reference
frame.
Let, the velocity $v$ be transverse to the electromagnetic wave,
e.g. the velocity $v$ be in the $z$ direction.
Due to the velocity of the frame
an additional magnetic field in the $x$ direction arises
\begin{equation}
\Delta H_x=-\frac{v}{c}E_y.
\label{eq:dh}
\end{equation}
This means the first order phase oscillations
\begin{equation}
\omega t-kr\propto1-\frac{v}{c}\cos(\omega t-kr).
\label{eq:h}
\end{equation}
The first order effect average for the period is equal to zero.

The magnetic field $\Delta H_x$ produces an additional
electric field in the $y$ direction
\begin{equation}
\Delta E_y=\frac{v}{c^*}\Delta H_x=\frac{v^2}{cc^*}E_y
\label{eq:de}
\end{equation}
where
\begin{equation}
c^*=c[1-\frac{v}{c}\cos(\omega t-kr)]
\label{eq:c}
\end{equation}
is the effective velocity of light given by the first order
effect.
This means the second order phase oscillations
\begin{equation}
\omega t-kr\propto1-\frac{v^2}{cc^*}\cos^2(\omega t-kr).
\label{eq:e}
\end{equation}
Due to the effective velocity of light the second order phase
oscillations yield the third order effect
average for the period $v^3/2c^3$.
An observer can detect such a phase shift of the third order with
the Michelson-Morley interferometer.

Observations~\cite{Lin} tell us that the earth moves with the
velocity, $v_{CMB}=369\ \mathrm{km/s}$,
in the direction,
$(\alpha,\delta)=(11.20^{h},-7.22^0)$,
with respect to CMB. Due to the effect under consideration
the velocity of the earth with respect to CMB can
be seen as a third order effect $v_{CMB}^3/2c^3$ in the
Michelson-Morley  experiment.
When changing the angle between $v_{CMB}$ and $c$, one can detect
the anisotropy of the phase shift in the Michelson-Morley
experiment.

M\"{u}ller et al.~\cite{Mu} reported the results of the modern
Michelson-Morley experiment.
Within the Robertson-Mansouri-Sexl test theory~\cite{MS},
they obtained the isotropy violation parameter,
$B=-2.2\pm 1.5\times 10^{-9}$.
The obtained
isotropy violation parameter corresponds to the phase shift,
$(-B/4)(1+\cos^2\chi)\cos^2\delta=8.4\pm 5.7\times 10^{-10}$,
where $\chi=42.3^0$ is the latitude of the set up,
$\delta=-7.22^0$ is the declination of $v_{CMB}$.
For the Michelson-Morley experiment, the effect under
consideration yields the maximum anisotropy of the phase shift,
$v_{CMB}^3/2c^3=9.3\times 10^{-10}$,
that is in agreement with the experimental value.

In summary, we have considered the phase shift of
the electromagnetic wave in the frame moving transverse to the
wave. Due to the velocity of the earth with respect to CMB the
effect under consideration yields the anisotropy of the phase
shift, $v_{CMB}^3/2c^3$, which can be seen
in the Michelson-Morley experiment. The predicted value is in
agreement with the anisotropy determined in the
Michelson-Morley experiment.

\end{document}